\documentclass[aps,pre,twocolumn,showpacs]{revtex4}
\usepackage{amsfonts,amssymb,amsmath,latexsym,epsfig} 
\begin{document}  

\title{Nontwist non-Hamiltonian systems}

\author{E. G. Altmann} 
\email{edugalt@pks.mpg.de}
\author{G. Cristadoro} 
\email{giampo@pks.mpg.de}
\author{D. Paz\'o} 
\email{pazo@pks.mpg.de}

\affiliation{Max Planck Institute for the Physics of Complex Systems, N\"othnitzer Stra{\ss}e 38, 01187 Dresden, Germany}

\date{\today}

\begin{abstract}

We show that the nontwist
phenomena previously  observed in Hamiltonian systems exist also in time-reversible
non-Hamiltonian systems. In particular, we study the two standard collision/reconnection
scenarios and we compute the parameter space breakup diagram of the shearless
torus. Besides the Hamiltonian routes, the breakup may occur due to the onset
of attractors.  We study these phenomena  in coupled phase oscillators and in non-area-preserving maps.

\end{abstract}


\maketitle

\section{Introduction}\label{sec.introduction}

Dynamical systems are usually divided in two 
classes, conservative and dissipative systems. 
Although this division is conceptually useful, it can be misleading as far as
there are systems that  display both dissipative and conservative
(quasi-Hamiltonian) dynamics, not only for different control parameters but
also coexisting in different regions of the phase space.  
Systems with this counterintuitive property are time-reversible~\cite{report}
but not Hamiltonian.
Two decades ago Politi {\em et al.}~\cite{politi86} reported on a system of
this type in a set of differential equations modeling a laser, 
and later other examples in arrays of Josephson junctions~\cite{tsang91,tsang91b} 
and coupled phase oscillators~\cite{topaj} have been investigated. 

Recently, many fundamental studies in Hamiltonian dynamics have focused on
{\em nontwist systems}, i.e., systems where the twist condition is violated. The twist
condition asserts the nondegeneracy of the frequencies in the integrable
regime and it is assumed in many fundamental mathematical results~\cite{arnold}. In 
generic situations we expect it to be locally violated. In such cases  it was shown
that  genuine
nontwist phenomena appear~\cite{howard,negrete}.
More recently, it has been shown that such nontwist effects are fundamental
for the understanding of many different physical systems (see
Ref.~\cite{morrison} and   references therein), e.g., the magnetic field lines
in reversed shear tokamaks~\cite{morrison,plasma,petrisor03} and the
zonal flows in geophysical fluid dynamics~\cite{negrete93}.

In this paper we provide  a novel link between these two  aspects
 of classical dynamics by showing that the nontwist phenomena, 
studied so far in Hamiltonian systems, occur also in time-reversible non-Hamiltonian systems. In
particular, we show the standard collision/reconnections around the shearless torus and we study
its breakup in a two-dimensional parameter space.
 Regarding the breakup of the shearless torus, we show that the usual Hamiltonian
routes as well as a ``dissipative route'' exist in time-reversible  non-Hamiltonian systems. 

The paper is organized as follows. In Sec.~\ref{sec.twist} we present the
definition of the twist condition and discuss the class of systems where 
it is expected to fail. In Secs.~\ref{sec.continuous}
and~\ref{sec.discrete} we present 
examples of the nontwist behavior in time-reversible non-Hamiltonian systems
in continuous and discrete time, respectively. We summarize our conclusions in
Sec.~\ref{sec.conclusion}.

\section{Twist condition in time-reversible systems}\label{sec.twist}

The study of near-integrable Hamiltonian systems has led to one of the greatest
successes of modern classical mechanics. Near-integrable Hamiltonian systems can be
written as~$H(\mathbf{I},\boldsymbol{\theta})=H_0(\mathbf{I})+\epsilon H_1(\mathbf{I},\boldsymbol{\theta}) $, where ~$(\mathbf{I},\boldsymbol{\theta})$
are the action-angle variables of the integrable 
Hamiltonian~$H_0$. In this approach, many
important mathematical results (e.g., the KAM theorem) are valid
assuming some nondegeneracy condition of the frequencies $\dot{\theta}_k=\partial
H_0(\mathbf{I}) /\partial I_k$ (see, e.g.,  Appendix 8 of Ref.~\cite{arnold}).  The
most common nondegeneracy condition is the twist condition, defined as 
\begin{equation}\label{eq.twist}
 \det \left|\frac{\partial
  \dot{\theta_k}}{\partial I_j} \right|\neq0 \;\;\text{and}\;\;\det \left| \frac{\partial
  \theta^{(k)}_{n+1}}{\partial I^{(j)}_n}\right| \neq 0, 
\end{equation}
for continuous and discrete time systems respectively. While the
continuous version needs some modification when the dynamics is  restricted 
to a specific energy shell ~\cite{arnold}, the discrete version can be directly
applied to both maps and  the reduced dynamics obtained by a Poincar\'e section of the flow.
In the (near-)integrable regime the twist condition assures that the frequency of the
invariant tori varies monotonically with the action. Conversely, a local violation
  of the  twist condition  usually implies the existence of a torus with  maximum or
minimum frequency, the so-called {\em shearless torus}. Around the shearless
torus different  {\it nontwist phenomena}  were
discovered in Hamiltonian systems,
e.g., separatrix reconnections and island chains collisions~\cite{howard},
 manifold reconnections of hyperbolic points in the chaotic
regime~\cite{corso}, meandering~\cite{wurm}, and the fractality  of the shearless torus
at criticality~\cite{negrete,negrete97,shinohara}. These phenomena were
 observed in area preserving maps~\cite{howard,negrete,saito} and Hamiltonian flows
 with one and a half~\cite{negrete93,shuckburgh} and two~\cite{stagika} degrees of freedom and 
they typically have a strong impact in transport properties of the system. 

In this paper we show the existence of the nontwist phenomena in
 time-reversible non-Hamiltonian maps and flows. A dynamical system is called time-reversible if
there is an involution $G$ (i.e., $G^2= Id$) that reverses the direction of time. For example
for dynamical systems described by a first order differential equation
$d\mathbf{x}/dt=\mathbf{F}(\mathbf{x})$ or by a mapping
$\mathbf{x}_{n+1}=L\mathbf{x}_{n}$  reversibility implies~\cite{report}
\begin{eqnarray}\label{invo}
\frac{d (G\mathbf{x})}{dt}= -\mathbf{F}(G(\mathbf{x})) \quad \text{and} \quad L \circ G \mathbf{x}_{n+1}= G \mathbf{x}_n,
\end{eqnarray}
respectively.
This condition alone does not ensure the observation of quasi-Hamiltonian
dynamics. Additionally, the dimension of the invariant set of $G$ should be large enough compared to the dimension $D$
of the phase space.
As discussed by Topaj and Pikovsky~\cite{topaj} $\mathrm{dim}(\mathrm{Fix}(G))\ge D/2$ ($D$ even), $(D-1)/2$ ($D$ odd) 
is usually required for continuous systems.

The twist condition is defined for time-reversible
non-Hamiltonian systems locally in the regions of the phase space where quasi-Hamiltonian
dynamics is observed. Similar to the Hamiltonian case, a
regime where the dynamics is locally integrable is needed in order to define 
action-angle variables~($I,\theta$) to be used in Eq.~(\ref{eq.twist}). Many
fundamental results have been obtained for the quasi-Hamiltonian dynamics of time-reversible
non-Hamiltonian systems~\cite{politi86} (including the extension of the KAM theorem
with the same optimal nondegeneracy conditions~\cite{sevryuk}), but until now
there was no systematic study of the breakdown of the twist condition in these
systems. 

\begin{figure}[!ht]
\centerline{
\includegraphics[width=\columnwidth]{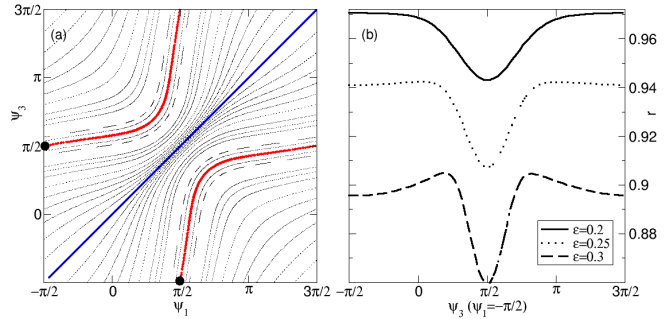}}
\caption{(Color online) (a) Poincar\'e section ($\psi_2=\pi/2$) of the 
system in Eq.~(\ref{eq.topaj}) for
  $\omega=1,\varepsilon=0.2$ where two nontwist tori are emphasized. IPs are
  marked with the symbol~$\bullet$. (b)
  Rotation number of the tori as a function of the coordinate $\psi_3$ at fixed
  $\psi_1=-\pi/2$ for $\omega=1$ and different values of $\varepsilon$ (see legend).} 
\label{fig1}
\end{figure}

A major question in this context is: in which kind of systems (Hamiltonian or time-reversible
non-Hamiltonian) the violation of the twist condition is to be expected?
Though we do not intend to give a general  answer to this question, we note
that systems possessing a symmetry in the phase space, apart from time
reversibility, exhibit naturally a shearless torus. Considering that two tori
related by this symmetry have the same frequency, and assuming
continuity, we conclude that  in the integrable limit there is at least one shearless torus between
 the two symmetric tori. This torus is invariant under the symmetry, what can be used
(together with the involution~$G$) to locate indicator points
(IPs)~\cite{shinohara}. IPs are  points that belong to
the shearless torus whenever it is not broken. 
Further examples are dynamical systems described by phase variables
(i.e., evolving on $\mathbb{T}^D$), where the symmetry is given by the
periodicity~$\phi=\phi+2\pi$ of the variables. This case is exemplified in
Fig.~\ref{fig1} for a non-Hamiltonian time-reversible 
system composed of phase oscillators, described by Eq.~(\ref{eq.topaj}) below. 
Fig.~\ref{fig1}(a) shows that the phase
space of the system is foliated by tori. Fig.~\ref{fig1}(b) shows the
rotation number of these tori as a function of one phase space variable for three
different control 
parameters~$\varepsilon$. Due to the periodicity of the
phase variables, at least two shearless torus exist (maxima and minima
of the frequency). As emphasized in Fig.~\ref{fig1}, one is located at the
diagonal and the second contains the IPs~$\psi_1=-\psi_3=\pm \pi/2$.

\section{Continuous-time systems}\label{sec.continuous}

Time-reversible non-Hamiltonian flows are encountered
in several physical situations (see \cite{report}
for a survey) such as
an externally injected class-B laser \cite{politi86} or 
arrays of $N$ coupled differential equations \cite{tsang91,topaj}.
One system where the properties discussed in Sec.~\ref{sec.twist} can be found is 
an array of $N$ coupled phase oscillators like the one 
 considered by Topaj and Pikovsky \cite{topaj}
 \begin{equation}
\dot {\varphi_k}=\Omega_k + \varepsilon f(\varphi_{k-1}-\varphi_k)+ \varepsilon f(\varphi_{k+1}-\varphi_k), \;\; k=1,...,N
\end{equation} 
with boundary conditions $\varphi_0=\varphi_1$, $\varphi_{N+1}=\varphi_N$ and 
vanishing coupling when the phases of the oscillators are equal, i.e., $f(0)=0$.
Taking the phase differences $\psi_k=\varphi_{k+1}-\varphi_k$ the number of variables
is reduced by one:
\begin{equation}
\dot{\psi_k}=\Delta_k+\varepsilon f(\psi_{k-1})+\varepsilon
f(\psi_{k+1})-2\varepsilon f(\psi_k), \;\; k=1,...,N-1
\label{psi}
\end{equation}
with $\Delta_k=\Omega_{k+1}-\Omega_k$.

If $f$ is an odd function and the natural frequencies are taken symmetrically $\Delta_k=\Delta_{N-k}$ the system
is reversible. The associated involution (\ref{invo}) is $G: \psi_k \rightarrow \pi - \psi_{N-k}$; 
and $\mathrm{Fix}(G)$ is given by $\psi_k+\psi_{N-k}=\pi$. 
In order to visualize the nontwist phenomena mentioned in Sec.~II
we restrict ourselves to three variables [$N=4$ in Eq.~(\ref{psi})], and we analyze 
the dynamics by means of a two-dimensional Poincar\'e section at~$\psi_2=\pi/2$.
The set of differential equations [for concreteness we already take $f(\cdot)=\sin(\cdot)$] 
we will study is:
\begin{equation}\label{eq.topaj}
\begin{array}{lll}
\dot{\psi_1}&=\omega-2 \varepsilon \sin\psi_1+\varepsilon \sin\psi_2,\\
\dot{\psi_2}&=1-2 \varepsilon \sin\psi_2+\varepsilon \sin\psi_1+\varepsilon \sin\psi_3,\\
\dot{\psi_3}&=\omega-2 \varepsilon \sin\psi_3+\varepsilon \sin\psi_2,
\end{array}
\end{equation}
where (rescaling coupling and time) we may set $\Delta_2=1$. In addition to the 
coupling variable $\varepsilon$, we take $\omega \equiv \Delta_{1}=\Delta_{3}$ as a second parameter of our
system. This allows a better exploration of the behaviors of the systems (e.g., codimension-two points); 
and it is also useful in order to compare our results to those obtained in 
the so-called standard nontwist map
by two-parameter sweep \cite{shinohara,wurm}.

Figure \ref{fig1}(a) shows a Poincar\'e section  of the phase space 
for small coupling $\varepsilon$. Phase space is foliated by tori with different rotation
numbers $\mathrm{r}$ [Fig.~\ref{fig1}(b)].
Equation (\ref{eq.topaj}) is invariant under the transformation $\psi_{1,3}\rightarrow\psi_{3,1}$
which imposes the existence of the invariant manifold $\psi_1=\psi_3$ that corresponds to
the diagonal shearless torus in Fig.~\ref{fig1}
The symmetry with respect to this diagonal torus together with the involution $G$ that maps
a torus into itself
allow us to locate the IPs ($\psi_1=-\psi_3=\pm \pi/2$) for the off-diagonal shearless torus
showed in Fig.~\ref{fig1}
[we will refer to it as {\em the} shearless torus (ST)].  
For the case $\varepsilon=0.2$ illustrated in
Fig.~\ref{fig1}(a) there are no further shearless tori. However, increasing
the control parameter the rotation number of these tori may pass from maximum to minimum (or 
  vice-versa) giving birth to other pairs of shearless
tori [e.g., $\varepsilon=0.3$ shown in Fig.~\ref{fig1}(b)]. We are specially interested  in
the region around the ST, where  we will report in what
follows the nontwist phenomena mentioned in Sec. II.    

\subsection{Collision/Reconnection Scenarios}

\begin{figure}[!ht]
\centerline{
\includegraphics[width=\columnwidth]{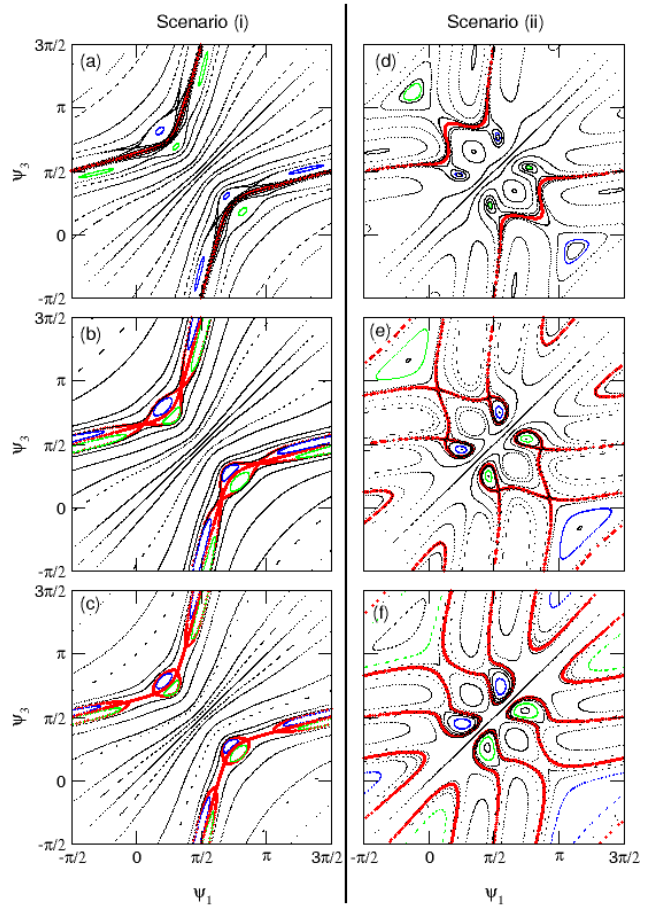}}
\caption{(Color online) Poincar\'e section of system~(\ref{eq.topaj}) for fixed
  $\varepsilon=0.25$ and different values of $\omega$. Sequence (a)
  $\omega=0.868$, (b) $\omega=0.8687606$, and (c) $\omega=0.869$ illustrates
  the collision of 3:4 island chains. Sequence (d)
  $\omega=0.801$, (e) $\omega=0.801523$, and (f) $\omega=0.802$
  illustrate a reconnection around 2:3 resonances.} 
\label{fig2}
\end{figure}

We start describing the two standard collision/reconnection scenarios studied
so far in nontwist Hamiltonian systems~\cite{howard}.
Figure \ref{fig2} shows the two scenarios for our system: 
periodic orbit collision (a-c) and separatrix reconnection (d-f).
We find the same phenomena as in Hamiltonian systems, the only appreciable
difference being the local non-volume preservation of the
flow~(\ref{eq.topaj}). Since in the quasi-Hamiltonian dynamics the phase-space
volume is preserved only in time-average, we see in Figs.~\ref{fig1}(a)
and \ref{fig2} bunches of tori compressed and expanded in different regions
of the phase space and Poincar\'e-Birkoff chains  composed of islands of
different sizes. Apart from these nonsymplectic features, Fig.~\ref{fig2}
essentially displays the two standard scenarios:  
\begin{enumerate}
\item[(i)]  Figure \ref{fig2}(a-c) shows the collision sequence of two symmetry related  Poincar\'e-Birkhoff chains with 
3:4 rotation number, by increasing $\omega$ at fixed $\varepsilon$. 
The collision occurs in Fig.~2(b), and in Fig.~2(c) one may see
the resulting dipolar structures formed by two saddle cycles 
on the torus and one center at each side of the ST. When $\omega$ is increased further 
the dipolar structure shrinks and finally disappears (at the point 
where the two saddles cycles on the torus annihilate each other).
\item[(ii)]  Figure \ref{fig2}(d-e) shows the reconnection scenario.
Twin Poincar\'e-Birkhoff chains with 2:3 rotation number are out of phase [Fig.~\ref{fig2}(d)]
and approach each other with the ST between them. At a critical
parameter~\footnote{Actually, the reconnection will 
not occur exactly in a line in the two-parameter space but in a (very) thin region as discussed in~\cite{comment}.} the separatrices of
both islands reconnect [Fig.~\ref{fig2}(e)]. Finally, passed the
 bifurcation, the torus exhibits a characteristic meandering,
due to the exchange of saddles between the two involved chains of islands.
\end{enumerate}

As usually found in the literature, scenario (i) is observed 
for rotation numbers with  even denominators, whereas scenario (ii)
applies to odd denominators. Nevertheless, 
as demonstrated in~\cite{petrisor03}, scenario (i) is not expected 
in the absence of symmetries
[scenario (ii) is then observed for both even and odd cases] (see, e.g.,~\cite{stagika}). 
As final remarks we mention that the shearless torus on the diagonal also exhibits 
the collision scenario [type (i)]. Heuristically the reconnection scenario [type (ii)]
cannot occur because this torus is constrained to the diagonal and cannot meander.
Another remark is that for resonances with small denominator (e.g.,  1:1 and 1:2) a
more complicated reconnection scenario occur due  to the
interplay of simultaneous resonances of both shearless tori (the diagonal one and the ST).

\subsection{Destruction of the shearless torus}

Our next numerical experiment was to vary the two control parameters $(\varepsilon, \omega)$,
in order to find the region of parameter space where the ST exists,
and in this way understand the routes for the destruction of the ST.
Due to the rather difficult numerical implementation of our system (if compared to a two dimensional map),
we needed to find an efficient method to detect the existence of the ST for 
a large set of parameter values.
By far, the most efficient way we found
was to resort to the Slater's theorem~\cite{slater}. In short, we start an initial condition
at one IP and
we check whether the number of different recurrence 
times to a small box around this point is at most three (see Appendix for details).

\begin{figure}[!ht]
\centerline{
\includegraphics[width=\columnwidth]{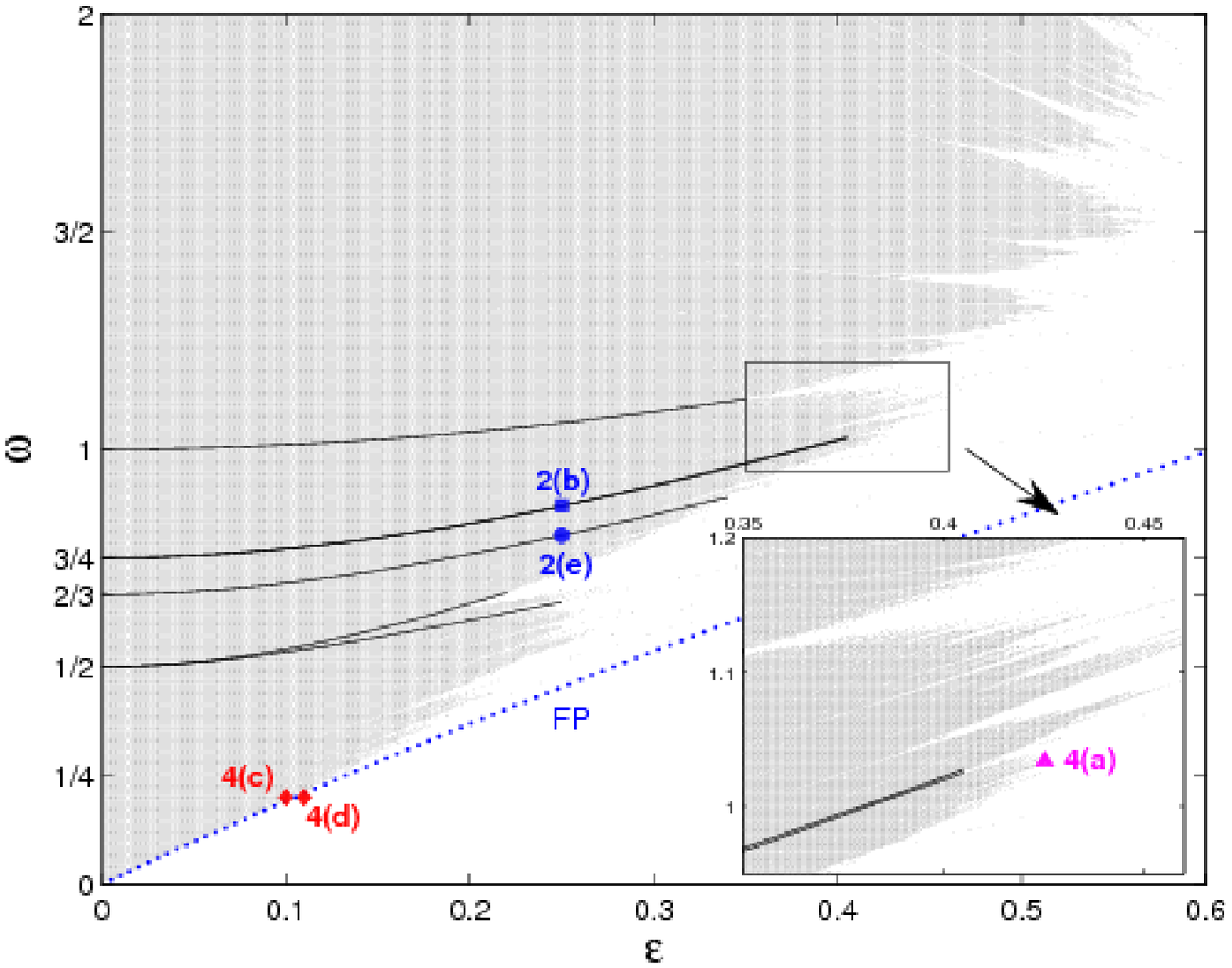}}
\caption{(Color online) Parameter space ($\varepsilon$,$\omega$) of Eq.~(\ref{eq.topaj}).
The shaded region indicates the parameters where the ST exists;
the inset evidences the fractal structure of the border.
Several solid lines indicate the collision/reconnection parameters for rotation numbers 1:2, 2:3, 3:4 and 1:1. 
The symbols ($\blacksquare, \bullet, \blacktriangle, \blacklozenge$)
are located at parameter values with characteristic dynamics shown in the corresponding figures.
The dotted line FP marks the onset of an attracting fixed point in the system.}\label{fig3}
\end{figure}

\begin{figure}[!ht]
\centerline{
\includegraphics[width=\columnwidth]{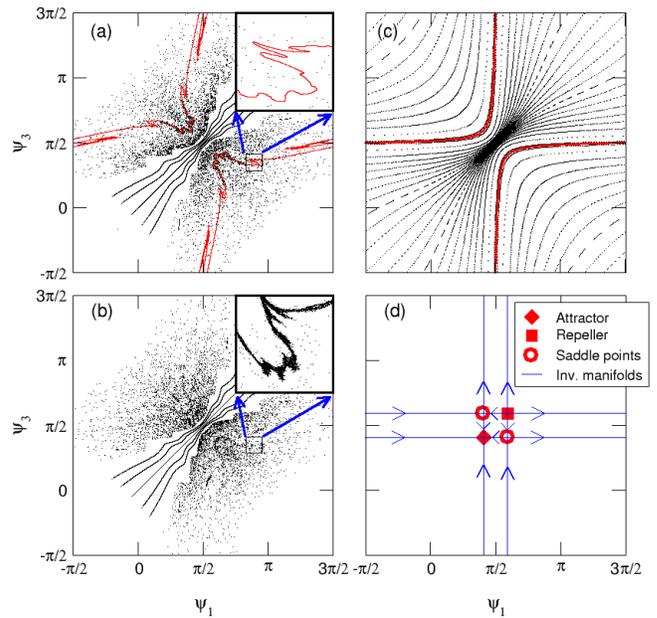}}
\caption{(Color online) Two different routes for the breakup of the
  shearless torus: 
  (a)-(b) Hamiltonian-like through critical point and (c)-(d) dissipative. (a)
  torus near criticality~$\varepsilon=0.425256,\omega=1.0335$ (the 
  inset shows the fractal structure of the torus). (b) Torus after criticality
  ~$\varepsilon=0.425257, \omega=1.0335$ (the inset show that the torus is
  destroyed). (c) Near-integrable phase space~$\varepsilon=0.1,
  \omega=0.2$. (d) Attracting fixed point~$\varepsilon=0.11,
  \omega=0.2$.}  
\label{fig4}
\end{figure}

The result of our calculation is shown in Fig.~\ref{fig3}. The shaded region indicates the 
parameter values where the ST exists, while in the white zone the ST does not exist.
The fractal character of the border limiting these two regions is apparent (see inset)
(cf.~\cite{shinohara,wurm}). The solid lines indicate the regions where a few collisions (1:2, 3:4) and reconnections (2:3,1:1) occur. 
Regions where  a couple of saddle orbits
lie on the ST [as in Fig.~\ref{fig2}(c)] are considered as regions where the ST exists 
(hence the shaded region between the two 1:2 lines in Fig.~\ref{fig3}).
As expected, the torus breakup is favored by the global bifurcations at the collisions/reconnections, and
accordingly there is a match between the solid lines and the white ``wedges'' in Fig.~\ref{fig3}. 
Singular points at the tips of the shaded region correspond to noble irrational rotation numbers 
where the torus
is expected to be fractal~\cite{negrete97,apte03,gaidashev}. Fig.~\ref{fig4}(a,b) show the torus
just before and after destruction.

Since the system is not Hamiltonian the existence of attractors cannot be excluded ~\cite{topaj}.
For instance, below the dotted line (denoted FP-line) in Fig.~(\ref{fig3}) a globally attracting fixed point exists (a
periodic orbit in the full phase space). 
Dissipation introduces  alternative ways for the destruction of the ST, as
shown in Fig.~\ref{fig4}(c,d). The bifurcation mechanism for the system~(\ref{eq.topaj}) follows:
At the  FP-line  a marginally stable fixed point is born on the invariant set of $G$
  at $\psi_1=\pi/2$ ($=\psi_3$). This point belongs to the diagonal shearless torus,
  which is broken only through this bifurcation. Below the FP-line four fixed points
exist: one attractor, one repeller (mirror of the attractor as demanded by
reversibility), and two saddles on the invariant set of $G$ [see
  Fig.~\ref{fig4}(d)]. 
Going back to the original system of $N=4$ phase oscillators, FP-line marks the onset
of synchronization of oscillators 1-2, and 3-4 (i.e.~a two-cluster state). Note that
the usual saddle-node bifurcation in this kind of transition to synchronization is forbidden due to
reversibility: for every attractor a mirror repeller must exist.
In a wide region above of the FP-line a (chaotic) attractor
and a set of Hamiltonian-like tori surrounding the diagonal torus coexist.

\subsection{Other systems}

In addition to the system studied above, Eq.~(\ref{eq.topaj}),
it is interesting to mention the case of an array of $N$ Josephson
junctions subject to a parallel resistive load.
This case may be described by {\em globally coupled}
phase oscillators (see \cite{tsang91} where the case $N=2$ is studied in detail). 
For $N=3$ and assuming that two of the junctions (say $j=1$ and $j=3$) are identical,
we get a set of equations with the same symmetries of Eq.~(\ref{eq.topaj})
 \begin{equation}\label{josephson}
 \begin{array}{lll}
 \dot{\phi_1}&=\Omega+ a \sin \phi_1 + \frac{1}{3} \sum_{j=1}^{3}  \sin\phi_j,\\
 \dot{\phi_2}&=\tilde \Omega+ \tilde a \sin \phi_2 + \frac{1}{3} \sum_{j=1}^{3}  \sin\phi_j,\\
 \dot{\phi_3}&=\Omega+ a \sin \phi_3 + \frac{1}{3} \sum_{j=1}^{3}  \sin\phi_j.
 \end{array}
 \end{equation}
We have observed in this system the same collision/reconnection scenarios described
previously for system~(\ref{eq.topaj}). As
before, the diagonal torus $\phi_1=\phi_3$ ceases to exist  by the
 onset of a stable fixed point (of the Poincar\'e section $\phi_2=\pi/2$). However, differently
from system~(\ref{eq.topaj}), at first this stable fixed point is not globally
attracting (as in the $N=2$ case studied in~\cite{tsang91}). Thus, the ``dissipative breakup'' of the
off-diagonal shearless torus does not occur right after the bifurcation but
only when the basin of attraction of the fixed point becomes the whole phase space.

\section{Discrete-time systems}\label{sec.discrete}

A simple procedure to obtain a time-reversible map is to integrate a time-reversible differential equation
$\dot{\mathbf{x}}=\mathbf{F}(\mathbf{x})$ semi-implicitly~\footnote{A.~Pikovsky,
private communication.}: 
\begin{equation}
\mathbf{x}_{n+1}-\mathbf{x}_{n}= k [\mathbf{F}(\mathbf{x}_{n+1})-\mathbf{F}(\mathbf{x}_{n})],
\label{implicit}
\end{equation}
where $k$ measures the finite-time integration step. 
Applying this procedure
to equations of phase oscillators, similar to those studied in
Sec.~\ref{sec.continuous}, we have obtained time-reversal maps on the torus 
where the same nontwist phenomena were observed. Certainly, the fact that mapping~(\ref{implicit})
is defined implicitly has several drawbacks, however, avoiding  details, we
note that the methods to construct  
(explicit) time-reversal mappings described in
Ref.~\cite{report} are not suited for the torus topology. 
In this section we study a different two-dimensional
time-reversible explicit map defined on a cylinder, which
illustrates that nontwist phenomena are not restricted to dynamics on a torus. 

Consider the following  map $L$
\begin{equation}\label{eq.map}
\begin{array}{ll}
y_{n+1} &= \dfrac{y_n+\text{a} \sin(2 \pi x_n)}{1+\text{b} y_n \sin(2 \pi x_{n})}, \\
x_{n+1} &= x_n+\cos(2 \pi y_{n+1})\;\;\; \text{mod(1)},
\end{array}
\end{equation}
which is a particular example of a class of  time-reversible maps discussed on
page 103 of Ref.~\cite{report}. 
Equation~(\ref{eq.map}) diverges for  $1+\text{b} y_n \sin(2 \pi x_{n})=0$ and thus
we restrict it  to the invariant region  $-\sqrt{\frac{\text{a}}{\text{b}}}<y<\sqrt{\frac{\text{a}}{\text{b}}}$
(limited by the invariant lines $y=\pm \sqrt{\frac{\text{a}}{\text{b}}}$) and the control
parameters to  $0 \leq \text{a} \text{b} < 1$.  

The map~(\ref{eq.map}) can be written as $L=M_1
\circ M_2$ , where
$M_{1,2}$ are the following involutions 
\begin{equation*}
\begin{array}{ll}
M_1:& x'=-x, y'=\dfrac{y+\text{a} \sin(2\pi x)}{1+\text{b} y \sin(2 \pi x)},\\ 
M_2:& x'=-x+\cos(2\pi y), y'=y.
\end{array}
\end{equation*}
 This property ensures  time
reversibility under $G=M_2$~\cite{report}. The determinant of the Jacobian is given
by $J=\frac{1-\text{a} \text{b} \sin^2(2 \pi x)}{[1+\text{b} y \sin(2\pi x)]^2}$ and shows 
that the map is Hamiltonian only in the case~$\text{b}=0$ (when one recovers the  Harper map~\cite{saito}). In the limit of small
control parameters the dynamics of map~(\ref{eq.map}) can be considered as a
perturbation of an integrable Hamiltonian system and it remains integrable almost
everywhere  since the KAM theorem, generalized to time-reversible systems, applies~\cite{sevryuk}.  

\begin{figure}[!ht]
\centerline{
\includegraphics[width=\columnwidth]{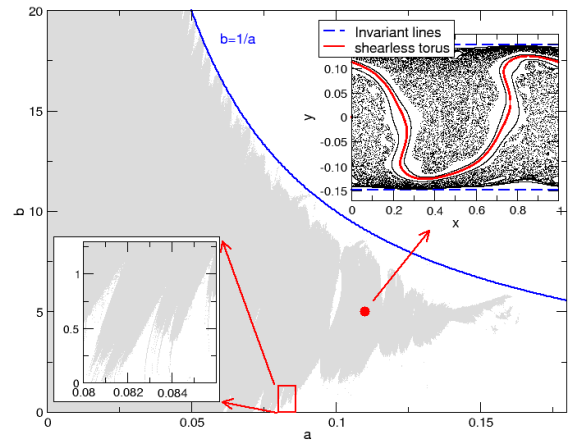}}
\caption{(Color online) Parameter space of the map~(\ref{eq.map}): 
  shaded regions indicate the parameters where the ST exists. Above the curve~$\text{b}=1/\text{a}$ the map is
  not defined. 
 Lower inset: magnification showing the white wedge structure
  in finer scale. Upper inset: phase space
  for  parameters~$\text{a}=0.11,\text{b}=5$. }
\label{fig5}
\end{figure}

In Sec.~\ref{sec.continuous} we argued that when at least one  symmetry is
present in the system, apart from time reversibility, the twist condition
must be violated in the near-integrable regime. In the case of 
map~(\ref{eq.map}) this symmetry is given by~$x'=x+\frac{1}{2}, y'= -y,$ 
which can be used (together with $M_2$) to determine the indicator points~$x=0.25,y=0$
and~$x=0.75,y=0$. According to Eq.~(\ref{eq.twist}) the twist condition is
violated in map~(\ref{eq.map})  at $y=-\text{a} \sin(2\pi x)$. In the upper
inset of Fig.~\ref{fig5} we show  
the phase space  of the map~(\ref{eq.map})
for~$\text{a}=0.11$, $\text{b}=5$. The ST intersects the curve where the
twist condition is violated~\cite{negrete}.  Using the method described in the Appendix we
determined the regions of the parameter space~$(\text{a},\text{b})$ where this torus
exists. The results are shown in Fig.~\ref{fig5} where a fractal-like border similar to
the one observed in Fig.~\ref{fig3} is clearly recognizable. Indeed, we have observed numerically that all nontwist
phenomena discussed in Sec.~\ref{sec.continuous} are also observed in the
time-reversible non-area-preserving map given by Eq.~(\ref{eq.map}).

\section{Conclusions}\label{sec.conclusion}

Time-reversibility does not appear exclusively in Hamiltonian
systems. Remarkably, there are time-reversible non-Hamiltonian systems that
may exhibit, in addition to dynamics approaching an attractor, quasi-Hamiltonian
dynamics. While major results in this context have been achieved in the past  twenty years, a 
complete understanding of which features of Hamiltonian dynamics can be extended to reversible systems is still lacking. 

Frequently, time-reversal symmetry is  not the only symmetry of a system, in which  
case we have argued that the twist condition (i.e.,~the nondegeneracy of the frequencies) 
is  violated. 
We have investigated the existence of nontwist phenomena, previously studied
in Hamiltonian systems, in time-reversible non-Hamiltonian systems. From our
results, obtained in both continuous-  and discrete-time systems, we may
conclude that the same nontwist phenomena are reproduced in
non-Hamiltonian systems. In order to study the breakup of the shearless torus  
we have developed a novel numerical method to compute the parameter space
breakup diagram (see Appendix).  Our method is potentially useful in a wide
class of problems concerning the detection of quasiperiodic motion in a
multiparameter space. We have identified the parameter regions of the usual
Hamiltonian routes of breakup of the shearless torus (through
collision/reconnection and through a critical-fractal torus) and the
parameters where the breakup is due to the onset of attractors.

Besides the theoretical interest of expanding the class of systems where the
 nontwist phenomena occur, our results are relevant  
 for the comprehension of specific time-reversible non-Hamiltonian
 systems. Many dissipative systems posses an attractor of limit-cycle type
 where a marginally stable phase dynamics  exists. The interaction of such systems may lead to 
 quasi-Hamiltonian phase dynamics, 
 in which case nontwist phenomena should be expected. Usually a large number of oscillators  
 are coupled, what provides additional motivation for the extension of the
nontwist phenomenology to higher dimensions.
Indeed conservative dynamics
has  been already observed in phase models such as the dynamics of solitary
waves in  chains of dispersively coupled oscillators~\cite{rosenau}, and 
the finite-dimensional Kuramoto model~\cite{popovych}. 
The role of time-reversibility and the violation of the twist condition in these
non-Hamiltonian models remains to be  understood.

\acknowledgements
We thank R.D. Vilela and J.S.E. Portela for a careful reading of the
manuscript. E.G.A. was supported by CAPES (Brazil).

\section*{Appendix} 

The knowledge of the critical parameter  values of a map where an invariant rotational
circle  breaks can be crucial in many situations.  We briefly describe here the
techniques previously employed to calculate the breakup diagram of the
shearless torus and we introduce a new method which has proven to be
computationally efficient. Even though we have applied our method to 
compute the breakup diagram  it is rather general and we believe that it can
 be useful whenever quasi-periodic motion has to be efficiently detected.

The first breakup diagrams were calculated~\cite{shinohara} 
considering the torus to be destroyed whenever a trajectory started at an IP leaves a region that
certainly  would contain the curve.  A slightly different method was proposed in  Ref.~\cite{wurm}  using
the fact that  a trajectory in a periodic or quasi-periodic motion leads to a converging  winding number.
 These procedures can be computationally expensive: in a typical mixed phase
 space we expect that a trajectory will stick around the complex structure of
 cantori and regular islands while wandering into the chaotic sea and the time
 needed  to detect the non-existence of the tori could be  very large. 
A refined method relies on Greene's residue criterium  and can be used when high accuracy is needed (e.g., to precisely locate critical points in
the breakup diagram corresponding to noble rotation numbers) while it is not suited to explore a large parameter region~\cite{apte03,negrete}.

Here we take advantage of a simple property of rotations in one dimension  to
develop a method that is together fast and quite robust, and hence suited to
scan a large set of parameters.
The quasi-periodic dynamics restricted to the torus  can be reduced to a simple rotation of the circle 
using a natural parameterization of the  curve (an example can be found
in~\cite{apte}).
Slater's theorem \cite{slater} states that for any irrational rotation
$\omega$ and for any connected interval there are at most three different
return times.  
Moreover, in the case of three different return times one of them is the sum of the 
other two and two of them are always consecutive denominators in the continued fraction  expansion of the irrational number $\omega$.

Accordingly, our method simply consists in counting the number of different
return times of the iterates of the IP $x_0$ inside an arbitrary region that
contains a connected part of the torus around $x_0$ (for the specific
cases Figs.~\ref{fig3} and \ref{fig5} we used a box centered at one
IP~\footnote{The shearless torus near criticality is strongly
  meandering and some care 
in the choice of the recurrence region  has to  be taken in order to avoid the
torus to multiple cross it.}). The torus
 is considered to be broken whenever  the different return times  violate the
 conditions imposed by Slater's theorem (e.g., whenever their number exceeds
 three). 
When the rotation number of the torus is not known a priori (as in our case)
the additional restriction that uses its continued fraction expansion can not
be used. 

The implementation of the procedure is straightforward provided
one point on the torus is known (IP). It is evident that the constraints on
the return times are quite restrictive and they are typically rapidly failing
when the torus is broken and the trajectory enters 
the chaotic sea.

\end{document}